\documentclass{article}

\usepackage{PRIMEarxiv}

\usepackage[utf8]{inputenc} 
\usepackage[T1]{fontenc}    
\usepackage{hyperref}       
\usepackage{url}            
\usepackage{booktabs}       
\usepackage{amsfonts}       
\usepackage{nicefrac}       
\usepackage{microtype}      
\usepackage{lipsum}
\usepackage{fancyhdr}       
\usepackage{graphicx}       
\graphicspath{{media/}}     

\usepackage{amsmath,amssymb,amsfonts}
\usepackage{algorithmic}
\usepackage{textcomp}
\usepackage{xcolor}
\usepackage{makecell}
\usepackage{multirow}

\usepackage{bm}
\usepackage{adjustbox}

\title{GreenSaliency: A Lightweight and Efficient Image Saliency Detection Method}

\author{
  Zhanxuan Mei \\
  University of Southern California \\
  Los Angeles, USA\\
  \texttt{zhanxuan@usc.edu} \\
   \And
  Yun-Cheng Wang \\
  University of Southern California \\
  Los Angeles, USA\\
  \texttt{yunchenw@@usc.edu} \\
   \And
  C.-C. Jay Kuo \\
  University of Southern California \\
  Los Angeles, USA\\
  \texttt{cckuo@sipi.usc.edu} \\
}

\begin{document}
\maketitle

\begin{abstract}
Image saliency detection is crucial in understanding human gaze patterns from visual stimuli.
The escalating demand for research in image saliency detection is driven by the growing necessity to incorporate such techniques into various computer vision tasks and to understand human visual systems.
Many existing image saliency detection methods rely on deep neural networks (DNNs) to achieve good performance.
However, the high computational complexity associated with these approaches impedes their integration with other modules or deployment on resource-constrained platforms, such as mobile devices.
To address this need, we propose a novel image saliency detection method named GreenSaliency, which has a small model size, minimal carbon footprint, and low computational complexity. GreenSaliency can be a competitive alternative to the existing deep-learning-based (DL-based) image saliency detection methods with limited computation resources.
GreenSaliency comprises two primary steps: 1) multi-layer hybrid feature extraction and 2) multi-path saliency prediction.
Experimental results demonstrate that GreenSaliency achieves comparable performance to the state-of-the-art DL-based methods while possessing a considerably smaller model size and significantly reduced computational complexity.
\end{abstract}


\section{Introduction}\label{G_sec:introduction}

The attention mechanism within the human visual system indicates the regions humans are interested in within the observed scenes \cite{itti1998model}.
Such an attention mechanism is often studied through the analysis of human eye movements recorded via gaze-tracking technology during the presence of visual stimuli, such as images. 
These fixations, collected from eye-tracker data, indicate the most compelling locations within a scene. 
Typically, a saliency map is derived from fixation maps through convolution with a Gaussian kernel, enhancing the representation of salient locations. 
This saliency map, constructed at the pixel level, represents regions of attention within the stimuli.
Image saliency detection is used to detect the most informative and conspicuous fixations within visual scenes to emulate the attention mechanisms exhibited by human eyes.
Fixation maps, obtained through human subject studies, and their corresponding saliency maps generated from these fixation maps are commonly regarded as ground truths (GTs), utilized in the training and evaluating of image saliency detection models.
Image saliency detection research predominantly falls within two categories: 1) human eye fixation prediction, which involves the prediction of human gaze locations on images where attention is most concentrated \cite{che2019gaze}, and 2) salient object detection (SOD) \cite{zhang2019synthesizing}, which aims to identify salient object regions within an image. 
This paper focuses on the former, which predicts the human gaze from visual stimuli.

Comprehending human gaze patterns within visual stimuli is essential in modeling visual attention. 
Image saliency detection, serving as either a preparatory step or a guiding principle in image processing, facilitates the identification of regions likely to command initial human attention. This insight finds applications across diverse domains.
In saliency-driven perceptual image compression methodologies \cite{patel2021saliency}, saliency cues inform the allocation of coding resources, directing more bits towards salient regions while economizing on less crucial areas.
Recognizing the non-uniform distribution of visual attention across image regions, local image saliency detection has been integrated into no-reference image quality assessment techniques \cite{yang2019sgdnet} to capture spatial attention discrepancies, thereby enhancing quality prediction accuracy.
Furthermore, investigations into saliency within visual stimuli yield valuable insights into the cognitive mechanisms governing human visual processing and attentional allocation, thus informing strategies for data augmentation~\cite{uddin2020saliencymix} and enhancing model interpretability through saliency-guided training procedures \cite{ismail2021improving}.
Overall, image saliency detection plays a fundamental role in various visual processing applications and human perception applications. It is an essential area of research and development in computer vision and related fields.

During earlier epochs, conventional methodologies for image saliency detection primarily followed a bottom-up approach \cite{itti1998model, harel2006graph}, where the features are not required to be pre-computed, thus offering versatility across a wide range of applications.
Recent advancements have witnessed the increase of DL-based saliency detection methods \cite{pan2016shallow, cornia2018predicting}, which have demonstrated remarkable efficacy of feature representation capabilities of Convolutional Neural Networks (CNNs), driven by large-scale image saliency datasets \cite{judd2009learning, borji2015cat2000, jiang2015salicon}.
Given the limited volume of data within the saliency domain compared to some of the more prominent computer vision tasks, transfer learning emerges as a pivotal mechanism for enhancing image saliency detection. 
Drawing inspiration from the massive success of deep convolutional models such as VGGNet \cite{simonyan2014very} and ResNet \cite{he2016deep} in classification tasks, particularly on benchmarks like ImageNet \cite{deng2009imagenet}, researchers have leveraged transfer learning to introduce pre-trained features from large-scale and external datasets into the image saliency detection methods. As a result, these methods achieve superior performance due to the expressive image features.
Integrating these large pre-trained models into mobile or edge devices is economically burdensome due to their substantial computational demands and expansive model sizes.
Given that image saliency detection methods typically serve as preliminary stages in image processing pipelines, allocating such considerable computational resources may be deemed unwarranted.

To address these challenges, we propose a lightweight image saliency detection method called GreenSaliency in this work. GreenSaliency features a smaller model size and
lower computational complexity while achieving comparable performance against DL-based methods.
The processing pipeline of GreenSaliency contains two modules: 1) multi-layer hybrid feature extraction and 2) multi-path saliency prediction.
We conduct experiments on two popular image saliency datasets and show thatGreenSaliency can offer satisfactory performance in five evaluation metrics while demanding significantly small model sizes, short inference time, and low computational complexity.
This paper has the following two main contributions. 
\begin{itemize}
\item 
Introduction of a novel method for saliency detection termed GreenSaliency, which is characterized by a transparent and modularized design. This method features a feedforward training pipeline distinct from the utilization of DNNs. The pipeline encompasses multi-layer hybrid feature extraction and multi-path saliency prediction.
\item Execution of experiments on two distinct datasets to assess the predictive capabilities of GreenSaliency. 
The findings illustrate its superior performance compared to conventional methods and its competitive standing against early-stage DL-based methods.
Furthermore, GreenSaliency exhibits efficient prediction performance compared to state-of-the-art methods employing pre-trained networks on external datasets while necessitating a significantly smaller model size and reduced inference complexity.
\end{itemize}

\section{Related Work}\label{G_sec:related}

\subsection{Conventional Image Saliency Detection Methods}
In the early stage of image saliency detection research, conventional methodologies predominantly adhered to a bottom-up approach.
For example, in \cite{itti1998model}, a visual attention system was inspired by early primate visual systems' behavioral and neuronal paradigms.
This system combines multi-scale image features into a unified topographical saliency map, subsequently utilized by a dynamical neural network to sequentially prioritize attended locations based on decreasing saliency levels. 
This process, informed by inputs from early visual processes, effectively simulates bottom-up attention mechanisms observed in primates.
Another prominent bottom-up model is the Graph-Based Visual Saliency (GBVS) framework \cite{harel2006graph}, comprising two sequential steps: 1) the formation of activation maps on specific feature channels followed by their normalization to accentuate salient features and 2) enabling combination with other maps.
Conversely, SUN \cite{zhang2008sun} introduces a Bayesian framework where bottom-up saliency naturally emerges as the self-information of visual features. In contrast, overall saliency, incorporating top-down and bottom-up influences, arises as the pointwise mutual information between features and the target during target search tasks.
Furthermore, \cite{kienzle2006nonparametric} proposes an image saliency model derived directly from human eye movement data, characterized by a nonlinear mapping from image patches to real values. This model is trained to produce positive outputs for fixated regions and negative outputs for randomly selected image patches.

\subsection{DL-based Image Saliency Detection Methods}

In recent years, deep-learning-based methodologies for saliency detection \cite{pan2016shallow, cornia2018predicting} have achieved notable performance owing to the advanced feature representation capabilities inherent in CNNs.
An early work in this domain was the emergence of eDN \cite{vig2014large}, which introduced hierarchical feature learning principles to visual saliency.
Additionally, \cite{pan2016shallow} introduced the concept of training a shallow CNN architecture from scratch, marking the inception of end-to-end CNN models tailored explicitly for image saliency detection tasks.
However, a significant impediment encountered in leveraging deep learning frameworks for image saliency detection lies in the scarcity of available data, primarily stemming from the laborious and cost-intensive nature of collecting fixation data. 
Furthermore, the sensitivity of image saliency to image transformations poses another challenge, significantly constraining potential data augmentations \cite{che2019gaze}.
Consequently, transfer learning has emerged as a pivotal strategy for refining image saliency detection performance.
Inspired by the remarkable success of deep convolutional networks in classification tasks, particularly exemplified by benchmarks like ImageNet, most high-performing saliency models have embraced transfer learning, typically relying on pre-trained models on ImageNet as the foundation. 
Noteworthy milestones in this trajectory include the pioneering work of DeepGaze I \cite{kummerer2014deep}, which has since evolved into DeepGaze II \cite{kummerer2017understanding}, leveraging the VGG19 architecture. Following this paradigm, DeepGaze IIE \cite{linardos2021deepgaze} was proposed, demonstrating the attainment of robust confidence calibration on unseen datasets through the principled amalgamation of multiple backbone architectures.
Additionally, the Saliency Attentive Model (SAM) \cite{cornia2018predicting} introduced a novel Attentive Convolutional Long Short-Term Memory (LSTM) mechanism, sequentially directing attention to diverse spatial regions within a feature stack to enhance image saliency detection.
EML-NET \cite{jia2020eml} introduced a scalable approach for integrating multiple deep convolutional networks of varying complexities as encoders for saliency-relevant features.
Meanwhile, UNISAL \cite{droste2020unified} unified saliency prediction across both image and video modalities, leveraging the entirety of available saliency prediction datasets to enrich analysis outcomes.

Various models have been developed employing intricate, deep architectures or extending existing ones that have demonstrated efficacy in other domains. Nonetheless, all these models have uniformly embraced transfer learning, initializing their architectures by pre-training them on outside datasets.
For instance, SalGAN \cite{pan2017salgan} introduces a generative adversarial model tailored for image saliency detection, comprising two interconnected networks: one responsible for generating saliency maps from raw pixel data of input images. At the same time, the other discriminates between predicted saliency maps and ground truth.
Similarly, GazeGAN \cite{che2019gaze} leverages a modified U-Net architecture as its generator, amalgamating classic``skip connections'' with a novel ``center-surround connection'' (CSC) module.
Departing from conventional feedforward architectures for saliency prediction, FBNet \cite{ding2021fbnet} integrates feedback convolutional connections to establish links between high-level blocks and low-level layers, thereby enhancing feature representation. 
Subsequently, SalFBNet \cite{ding2022salfbnet} introduces a feedback-recursive convolutional framework. 
Additionally, \cite{ding2022salfbnet} pioneers creating a large-scale Pseudo-Saliency dataset, addressing concerns regarding data scarcity in image saliency detection.

\subsection{Green Machine Learning}

Recently, the concept of green learning~\cite{kuo2023green} has emerged as a novel machine learning paradigm to develop efficient models with a reduced carbon footprint.
These models are distinguished by their compact sizes and minimized computational complexities during the training and inference stages. 
An additional benefit lies in their mathematical transparency, facilitated by a modularized design principle.
In contrast to DNNs, which rely on backpropagation for iterative parameter updates, green learning adopts a one-pass training pipeline comprising three sequentially cascaded modules: 1) unsupervised representation learning, 2) supervised feature learning, and 3) supervised decision learning.
The unsupervised representation learning module employs a hierarchical approach utilizing multiple Saab transforms \cite{kuo2019interpretable, chen2020pixelhop++}, thereby broadening the spectrum of candidate representations to enhance the discriminant capabilities of the subsequent supervised feature learning module. 
Subsequently, the Relevant Feature Test (RFT) \cite{yang2022supervised} is employed to discern the most powerful features from a plethora of candidate representations obtained from the preceding module.
Finally, a regressor, such as XGBoost~\cite{chen2016xgboost}, is trained to map the feature space to the target label space. 
Green Learning techniques have found applications across diverse domains, including deepfake detection \cite{chen2021defakehop}, blind image quality assessment \cite{mei2022greenbiqa},
image generation \cite{lei2021tghop}, etc.
In this paper, we propose a lightweight saliency detection method inspired by the principles of green learning.

\begin{figure*}[!htbp]
\centering
\includegraphics[width=1\linewidth]{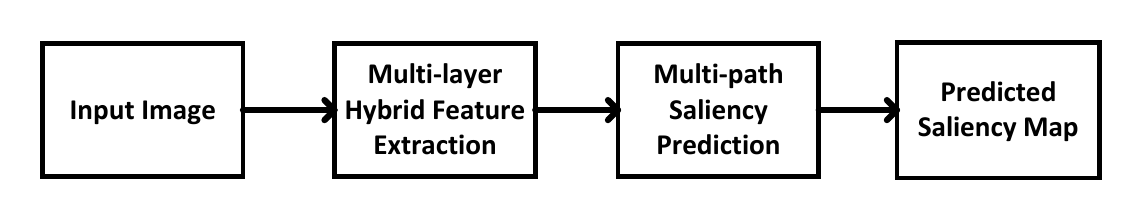}\\
\caption{An overview of the proposed GreenSaliency method.}\label{fig:pipeline}
\end{figure*}

\section{Proposed GreenSaliency Method}\label{G_sec:BIQA_method}

An overview of the proposed GreenSaliency method is depicted in Fig.
\ref{fig:pipeline}. As shown in the figure, GreenSaliency has a modularized
solution that consists of two modules: 1) multi-layer hybrid feature extraction and 2) multi-path saliency prediction. They are elaborated below.

\begin{figure*}[!htbp]
\centering
\includegraphics[width=0.7\linewidth]{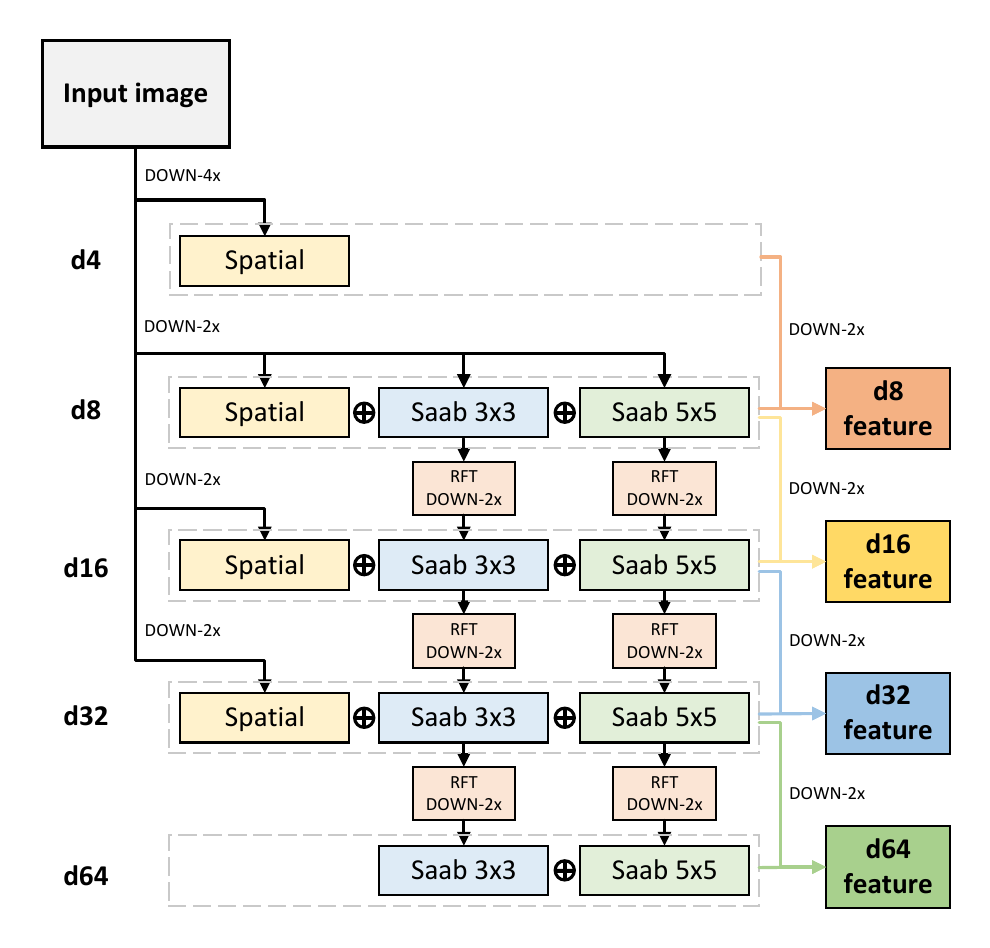}\\
\caption{Multi-layer hybrid feature extraction pipeline.}\label{fig:feature_extraction}
\end{figure*}

\subsection{Multi-layer Hybrid Feature Extraction} \label{subsec:feature_extraction}
Figure \ref{fig:feature_extraction} shows the multi-layer hybrid feature extraction pipeline.
In this section, we commence by delineating the multi-layer structure. 
Each layer comprises analogous components, such as spatial feature extraction and two Saab transforms, aimed at extracting hybrid features for each layer. Subsequently, the details of the spatial feature extraction module and the saab transform are introduced.

\subsubsection{Multi-layer Structure}

A hierarchical multi-layer structure is employed to capture features spanning from local to global contexts across different receptive fields. 
This structure comprises five layers denoted as d4, d8, d16, d32, and d64, respectively. Within each layer, except d4 and d64, a spatial feature extraction module accompanied by two Saab transform modules exists. 
The spatial feature extraction module facilitates the direct extraction of spatial features from input images that have undergone downsampling.
Notably, the prefix ``d'' within the naming of each layer signifies downsampling, wherein, for instance, d4 indicates that spatial features at this layer are computed based on input images downsampled by a factor of 4. 
The two Saab transform modules, employing kernel sizes of $3 \times 3$ and $5 \times 5$, respectively, compute Saab coefficients derived from those of the previous layer.
This recursive process enables extracting high-level features as the layers progress deeper.
Further elaboration on the spatial feature extraction module and Saab transform mechanisms is provided in Section \ref{subsubsec:spatial} and Section \ref{subsubsec:saab}, respectively.

The Relevant Feature Test (RFT) is employed to discern the most powerful coefficients or channels meriting propagation to the subsequent layer after obtaining Saab coefficients from the two Saab transform modules within the current layer. Section \ref{subsubsec:rft} provides a detailed exposition of the RFT methodology.
Additionally, the selected coefficients undergo downsampling at a ratio of 2 to mitigate the computational burden associated with processing coefficients in the subsequent layer.
Within each layer, the computed spatial features and Saab coefficients derived from the two Saab transform modules are concatenated to form hybrid features. 
Subsequently, these hybrid features are combined with hybrid features from other layers to compose sets of features utilized for image saliency prediction.
Illustrated in Figure \ref{fig:feature_extraction}, four sets of features are discernible, denoted as d8 features, d16 features, d32 features, and d64 features, respectively. 
Each set encompasses extracted features from two layers. For instance, d8 features comprise hybrid features derived from layers d4 and d8. 
These four sets of features are transmitted to the multi-path saliency prediction module, as elaborated in Section \ref{subsec:multi-path}, to predict the corresponding saliency map or residual.

\subsubsection{Spatial Feature Extraction}\label{subsubsec:spatial}

In contrast to features computed by the Saab transform, which are propagated from shallow layers to deeper layers, the spatial features extracted in each layer are calculated directly from down-sampled input images and are not transmitted to deeper layers.
The rationale behind extracting spatial features lies in their ability to encapsulate local and spatial information inherent in the input images. In contrast, features computed by the Saab transform primarily capture spectral information. 
Spatial features consist of three primary components: 1) local Saab features, 2) edge features, and 3) location features.

The local Saab features are derived by implementing two sequential cascade Saab transforms. Distinguished from the two Saab transforms within the same layer, these cascade Saab transforms are characterized by smaller kernel sizes in the spatial domain, specifically $2 \times 2$ and $3 \times 3$.
This distinction in kernel sizes is attributed to their primary focus on capturing and emphasizing local information within the directly downsampled input images.
Utilizing smaller kernel sizes enables these cascade Saab transforms to effectively highlight and extract intricate details locally, contributing to the overall feature representation and analysis.
Concerning edge features, they are derived utilizing the Canny edge detector \cite{canny1986computational}, a widely employed method for detecting edges in images.
Numerous prior investigations have established that objects near an image's center garner greater attention from observers \cite{judd2009learning}. 
This observation suggests that locations proximal to the image center are more likely to exhibit saliency than those farther away. A Gaussian distribution can effectively model this empirical observation.
Let $(c_x, c_y)$ represent the center coordinates of an image; subsequently, the location feature $f(x, y)$ at coordinates $(x, y)$ can be expressed as a Gaussian map defined by
\begin{equation}
f(x, y) = exp\left(-\frac{(c_x-x)^2+(c_y-y)^2}{\sigma^2}\right),
\end{equation}
where $\sigma$ is a pre-defined parameter.

\subsubsection{Subspace Approximation with Adjusted Bias (Saab) Transform}\label{subsubsec:saab}

The initial step in the processing pipeline involves partitioning the input images into overlapping blocks, typically of sizes $3 \times 3$ or $5 \times 5$, followed by applying the Saab transform \cite{kuo2019interpretable}.
The Saab transform, a principal component analysis (PCA) method with mean-removal, distinguishes itself by incorporating an additional bias vector.
Within the framework of the Saab transform, a constant-element kernel is utilized to compute the average value of image patches, commonly known as the DC (Direct Current) component.
Subsequently, PCA is applied to these patches post-removal of the computed mean, yielding data-driven AC (Alternating Current) kernels. Applying these AC kernels on individual patches leads to extracting AC coefficients associated with the Saab transform.
For instance, an input cuboid has dimensions $(H \times W) \times C$, where $H$, $W$, and $C$ denote the height, width, and number of channels, respectively.
To execute a $3 \times 3$ Saab transform, the input cuboid is initially subdivided into overlapping cuboids of size $(3 \times 3) \times C$, with a stride of 1 and padding applied as necessary.
Then, these cuboids are flattened into 1-dimensional vectors, each possessing a length of $9C$.
The mean values of these vectors are computed to yield the DC channel, while the mean-removed vectors undergo PCA transform, resulting in the generation of $9C - 1$ AC channels.
Following the computation of the $3 \times 3$ Saab transform, the input cuboid, originally of size $(H \times W) \times C$, transforms to dimensions $(H \times W) \times 9C$, preserving the spatial resolution and featuring $9C$ Saab coefficient channels.

In the feature extraction pipeline context, a multi-layer pipeline comprising four successive Saab transforms, located in d8, d16, d32, and d64, is adopted to decorate the DC coefficients and generate higher-level representations. 
This multi-layer pipeline facilitates the transformation of input data into a more informative and discriminative feature space, thereby enhancing the effectiveness of subsequent processing steps. 

\begin{figure*}[t]
\centering
\includegraphics[width=.3\linewidth]{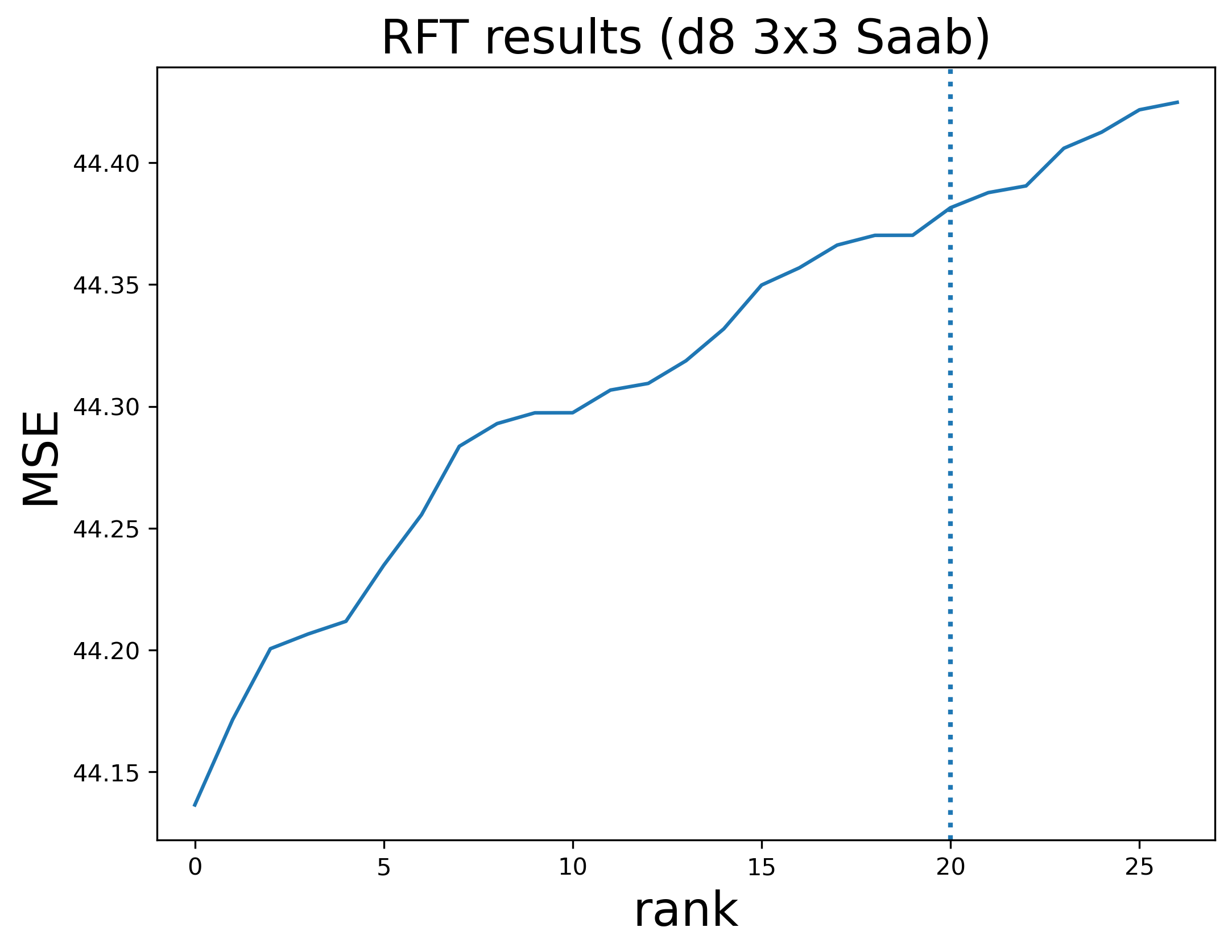} 
\includegraphics[width=.3\linewidth]{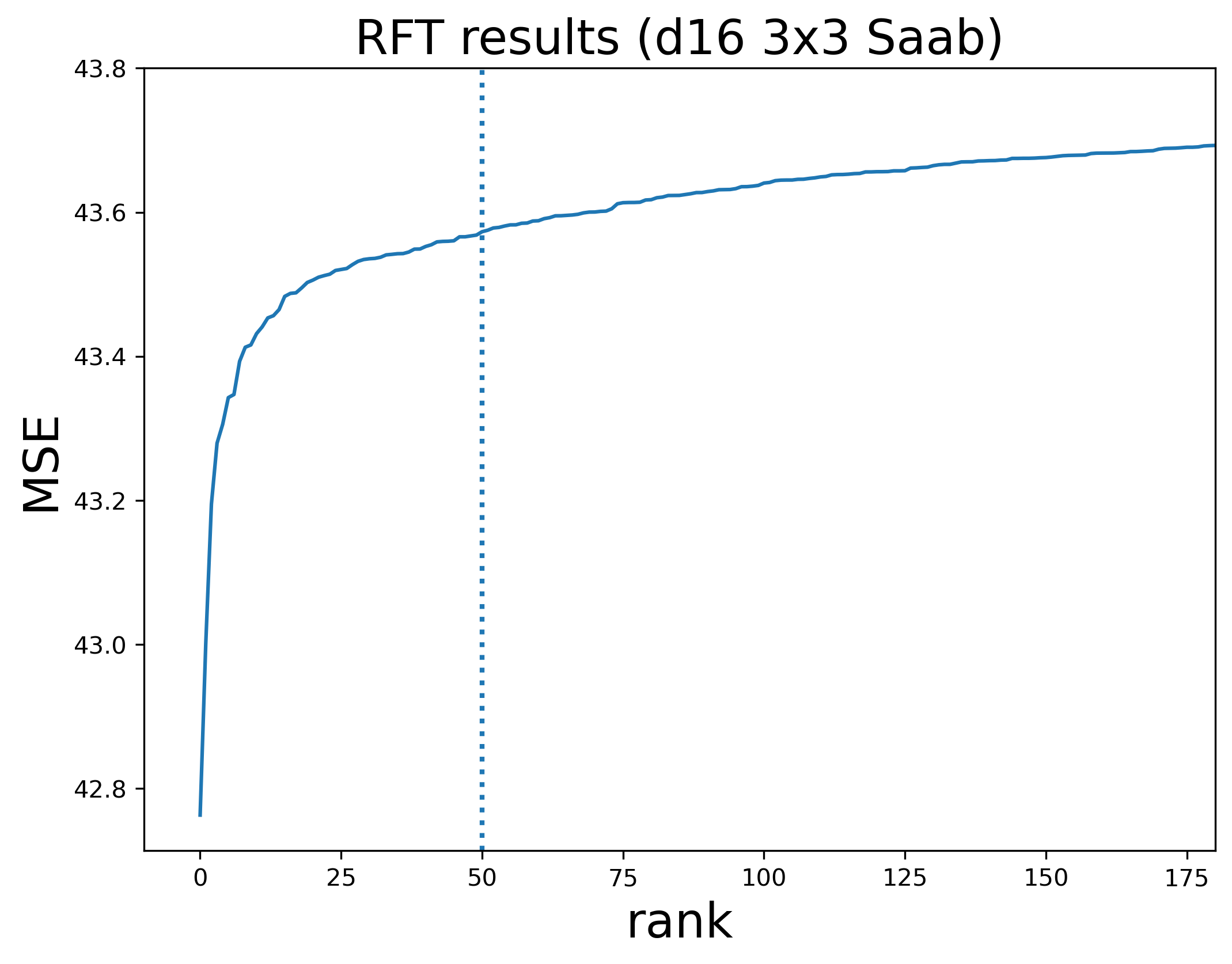} 
\includegraphics[width=.3\linewidth]{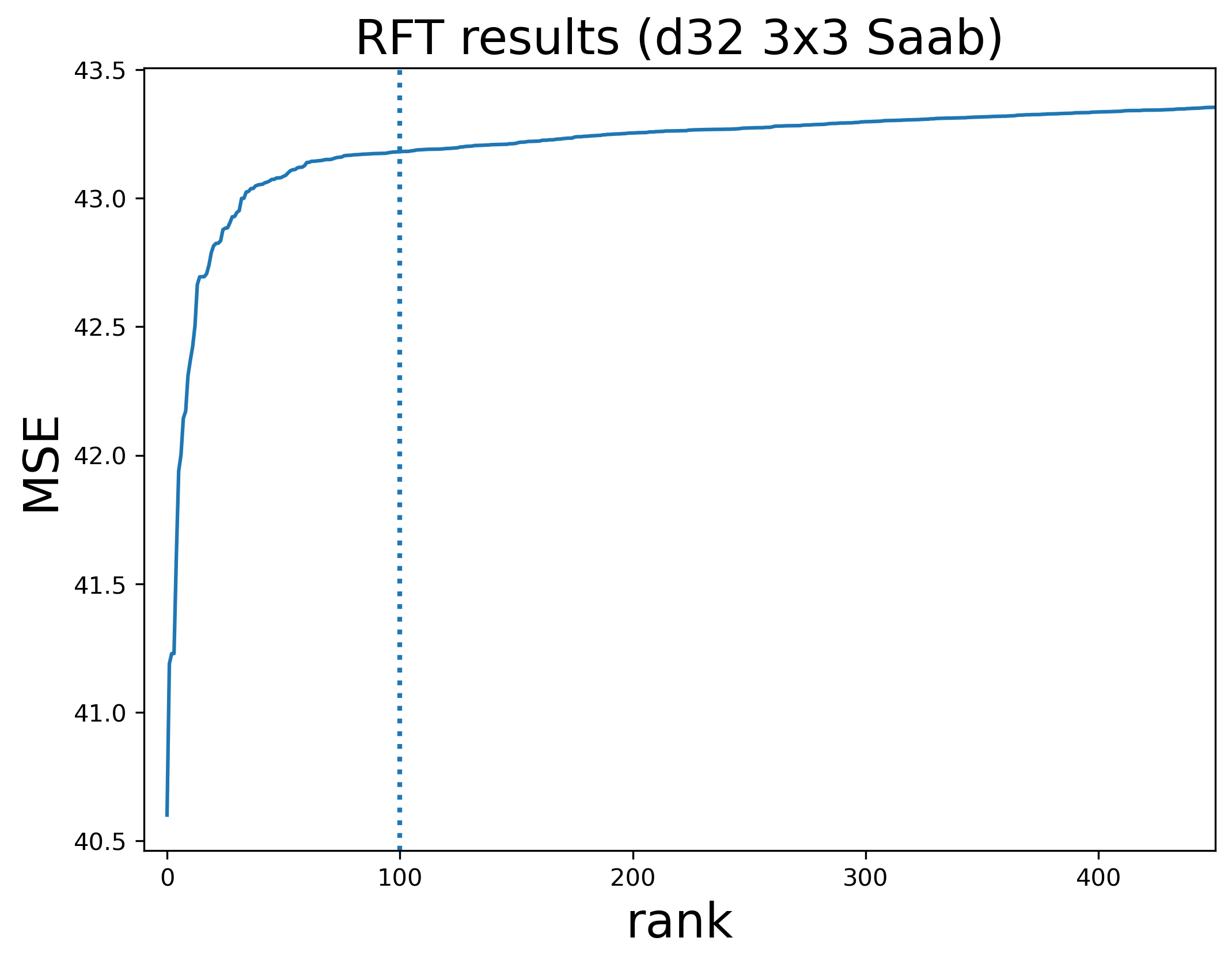} \\
\scriptsize(a) \hspace{3.1cm} \scriptsize(b) \hspace{3.1cm} \scriptsize(c) 
\caption{RFT results of $3 \times 3$ Saab coefficients from three layers: (a) d8, (b) d16, and (c) d32.}\label{fig:rft_3}
\end{figure*}

\subsubsection{Relevent Feature Test (RFT)}\label{subsubsec:rft}

The objective of feature selection is to identify highly discriminative features from a diverse candidate set of extracted features. To accomplish this task, a robust technique known as the Relevant Feature Test (RFT) \cite{yang2022supervised} is employed.
RFT entails dividing a feature dimension into two left and right segments and evaluating the total mean-squared error (MSE) between them.
The resultant approximation error is the RFT loss function, with a smaller RFT loss indicating a more informative feature dimension.
Given a dataset comprising $N$ data samples and $P$ features, each feature dimension, denoted by $f_i$ where $1 \leq i \leq P$, each feature dimension possesses a minimum and maximum range of $f^i_{min}$ and $f^{i}_{max}$, respectively. 
The deployment of RFT involves three distinct steps, delineated as follows.
\begin{itemize}
\item Training Sample Partitioning. 
The primary objective entails identifying the optimal threshold, $f^i_{op}$, within the range [$f^i_{min}$, $f^i_{max}$], facilitating the partitioning of training samples into two subsets: $S^i_L$ and $S^i_R$. 
If the value of the ith feature, $x^i_n$, for the $n$th training sample $x_n$ is lower than $f^i_{op}$, then $x_n$ is assigned to $S^i_L$; otherwise, $x_n$ is allocated to $S^i_R$. 
To refine the search space for $f^i_{op}$, the entire feature range, [$f^i_{min}$, $f^i_{max}$], is divided into $B$ uniform segments, and the optimal threshold is sought among $B-1$ candidates.

\item RFT Loss Measured by Estimated Regression MSE. Denoting the regression target value as $y$, $y^i_L$, and $y^i_R$ represent the mean target values in $S^i_L$ and $S^i_R$, respectively.
These mean values are the estimated regression values for all samples in $S^i_L$ and $S^i_R$.
The RFT loss is defined as the summation of the estimated regression MSEs of $S^i_L$ and $S^i_R$, given by
\begin{equation}
R^i_t = \frac{N^i_{L,t}R^i_{L,t}+N^i_{R,t}R^i_{R,t}}{N},
\end{equation}
where $N^i_{L,t}$, $N^i_{R,t}$, $R^i_{L,t}$, and $R^i_{R,t}$ represent
the sample numbers and estimated regression MSEs in subsets $S^i_L$
and $S^i_R$, respectively.  
Each feature $f^i$ is characterized by its
optimized estimated regression MSE over a set, $T$, of candidate
partition points:
\begin{equation}
R^i_{op} = \mathop{\min}_{t\in T} R^i_t.
\end{equation}

\item Feature Selection based on the Optimized Loss. The optimized estimated regression MSE value, $R^i_{op}$, is computed for each feature dimension, $f_i$.
These values are subsequently arranged in ascending order, reflecting the relevance of each feature dimension.
A lower $R^i_{op}$ value denotes a higher relevance of the $i$-th-dimensional feature, $f^i$.
\end{itemize}

After computing the $R^i_{op}$ value for each feature dimension, $f^i$, representation indices $i$ are arranged based on their MSE values in ascending order.
Within the multi-layer hybrid feature extraction pipeline context, RFT is employed between two consecutive layers to diminish the feature dimensionality from the shallow layer onwards.
The RFT outcomes of the $3 \times 3$ Saab coefficients computed in layers d8, d16, and d32 are shown in Figure \ref{fig:rft_3}.
In these figures, discernible elbow points are observed in the curves of the RFT outcomes, especially for the results from d16 and d32. 
The dot lines represent the number of selected features.
Consequently, we select the top-ranking features, representing the most influential ones, for propagation to deeper layers.
The primary rationale behind incorporating RFT between successive layers lies in the recognition that not all channels of coefficients necessitate processing at deeper layers, primarily due to the substantial computational complexity entailed.
Hence, RFT identifies the most pertinent coefficients or channels that warrant deeper layer-level processing.

\begin{figure*}[!htbp]
\centering
\includegraphics[width=1\linewidth]{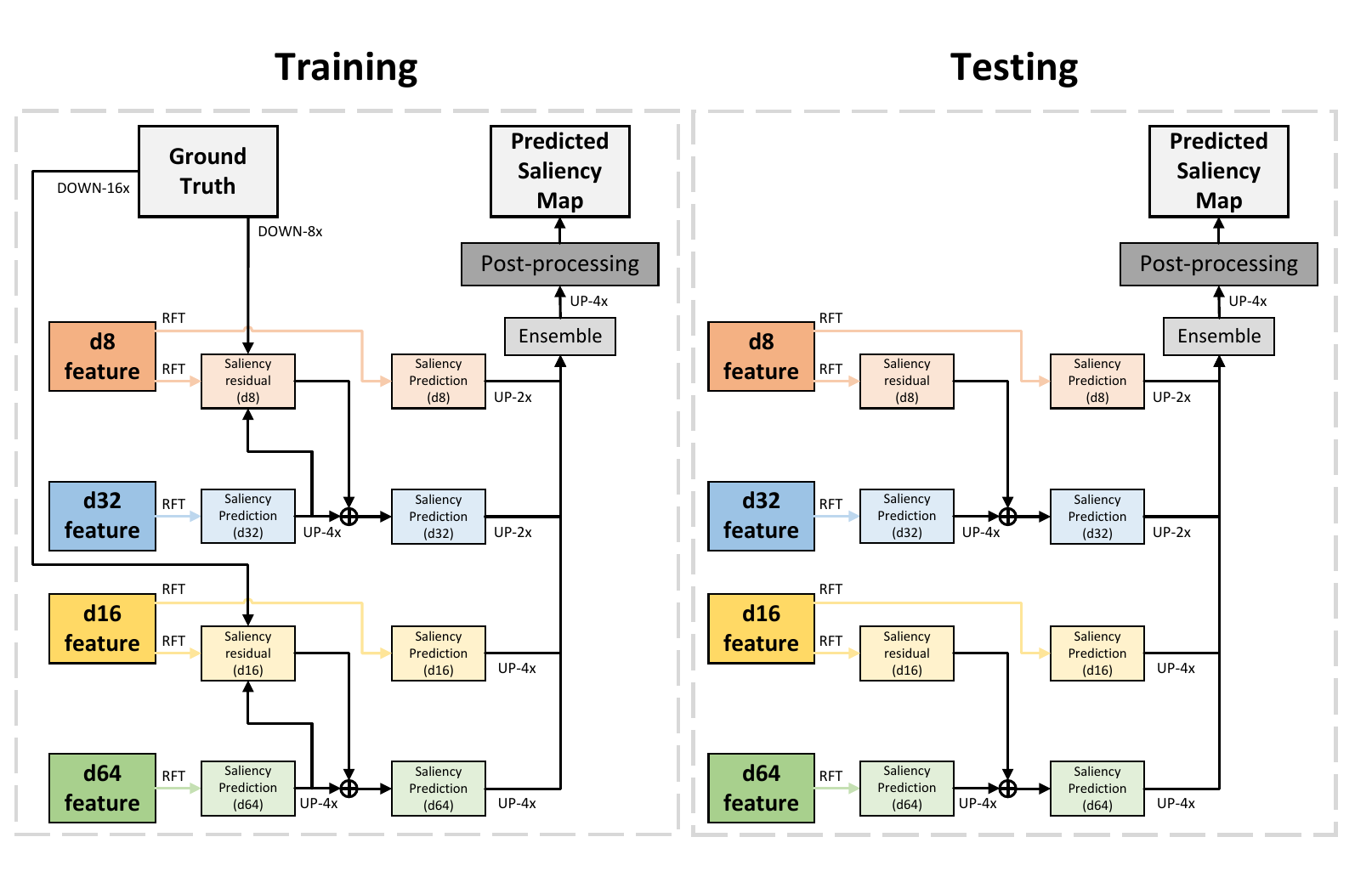}\\
\caption{Multi-path saliency prediction pipeline.}\label{fig:multi-path}
\end{figure*}

\begin{figure*}[!htbp]
\centering
\includegraphics[width=1.0\linewidth]{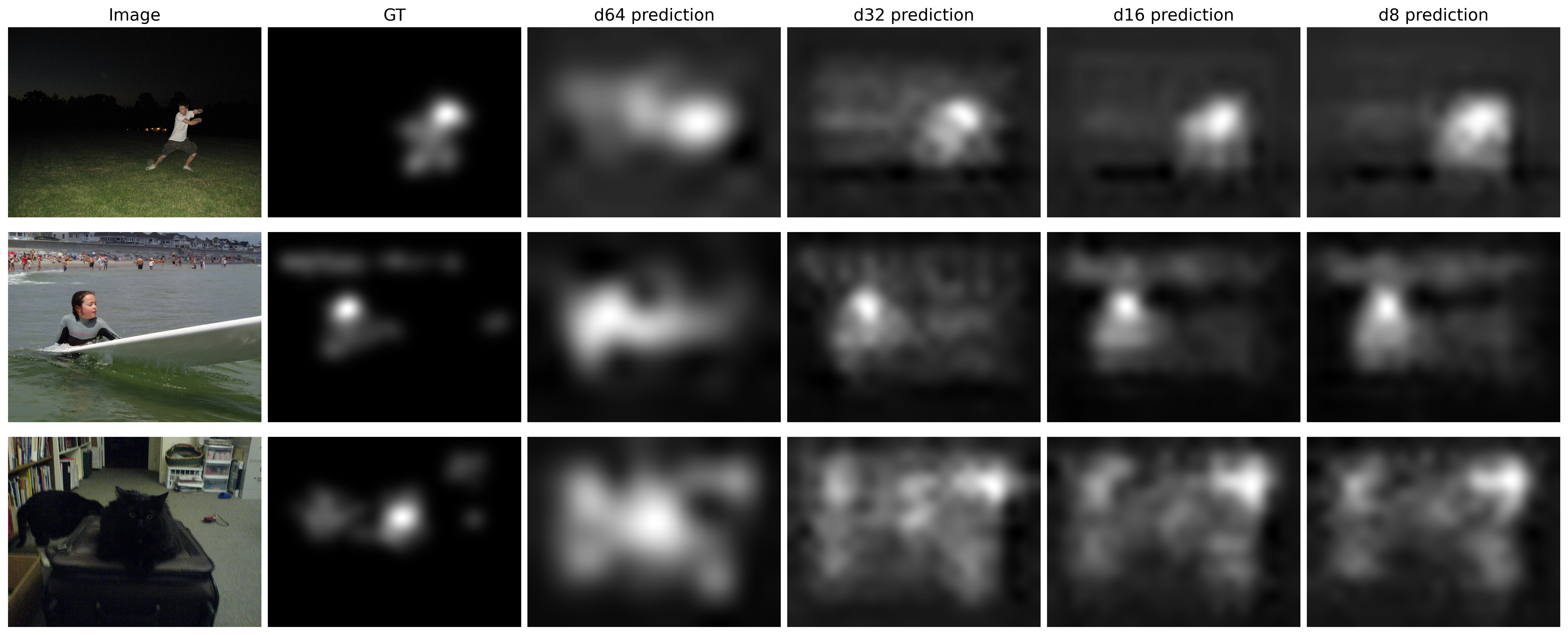}\\
\caption{Prediction results from different layers.}\label{fig:self_cases}
\end{figure*}

\subsection{Multi-path Saliency Prediction}\label{subsec:multi-path}

Following the extraction of features from the multi-layer hybrid feature extraction module, these features are utilized for multi-path saliency prediction, as delineated in Figure \ref{fig:multi-path}.
Four distinct paths are initiated, corresponding to four layers: d8, d16, d32, and d64, each tasked with predicting its corresponding saliency map.
Within each path, two conditions are considered: saliency map prediction and saliency residual prediction, whose details are expounded upon in Section \ref{subsubsec:residual}.
Upon obtaining predicted saliency maps from the four paths, they are aggregated within the ensemble module and subjected to post-processing to yield the final predicted saliency map. 
Detailed discussions on the ensembles and post-processing modules are provided in Section \ref{subsubsec:ensemble} and Section \ref{subsubsec:post}, respectively.

The rationale for adopting multi-path saliency prediction instead of a singular path employing all features stems from the recognition that features extracted from diverse layers encapsulate information from disparate receptive fields. 
Such diversity in receptive fields proves advantageous, catering to the varying characteristics of different input images.

In the context of the multi-layer Saab transform, the features utilized for predicting the saliency map in layers d64 and d32 are relatively high-level features, while those in layers d16 and d8 are low-level features.
High-level features, characterized by a large receptive field, concentrate on capturing the overall structure or shape of the entire saliency map rather than focusing on small objects or details. 
Conversely, low-level features exhibit greater sensitivity towards smaller objects while potentially overlooking the broader context of the scene.
Figure \ref{fig:self_cases} demonstrates the efficacy of utilizing distinct feature sets, where predicted saliency maps derived solely from specific feature sets are showcased.
In the first two rows, predictions based on d8 and d16 features outperform those based on d32 and d64 features, attributed to the high-level features in d32 and d64 inadequately capturing the relatively small human subjects in these images.
However, in the third row, high-level features excel in delineating the shape of the cat in the center. In contrast, predictions based on d8 and d16 features primarily focus on the top-right corner, illustrating the nuanced interplay between feature levels and their corresponding receptive fields.
Section \ref{subsec:ablation} illustrates more experimental results of different layers.

\begin{figure*}[t]
\centering
\includegraphics[width=.45\linewidth]{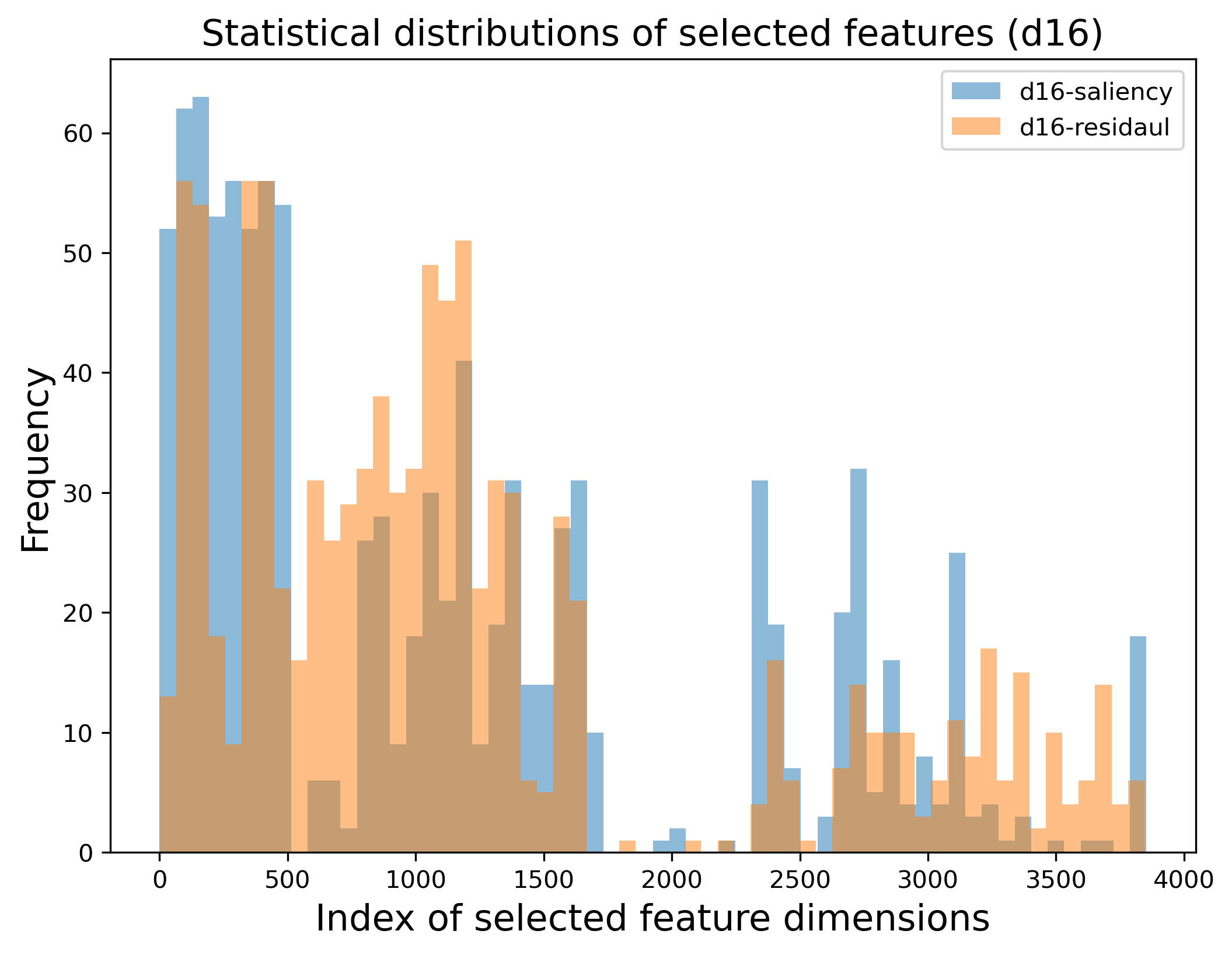} 
\includegraphics[width=.45\linewidth]{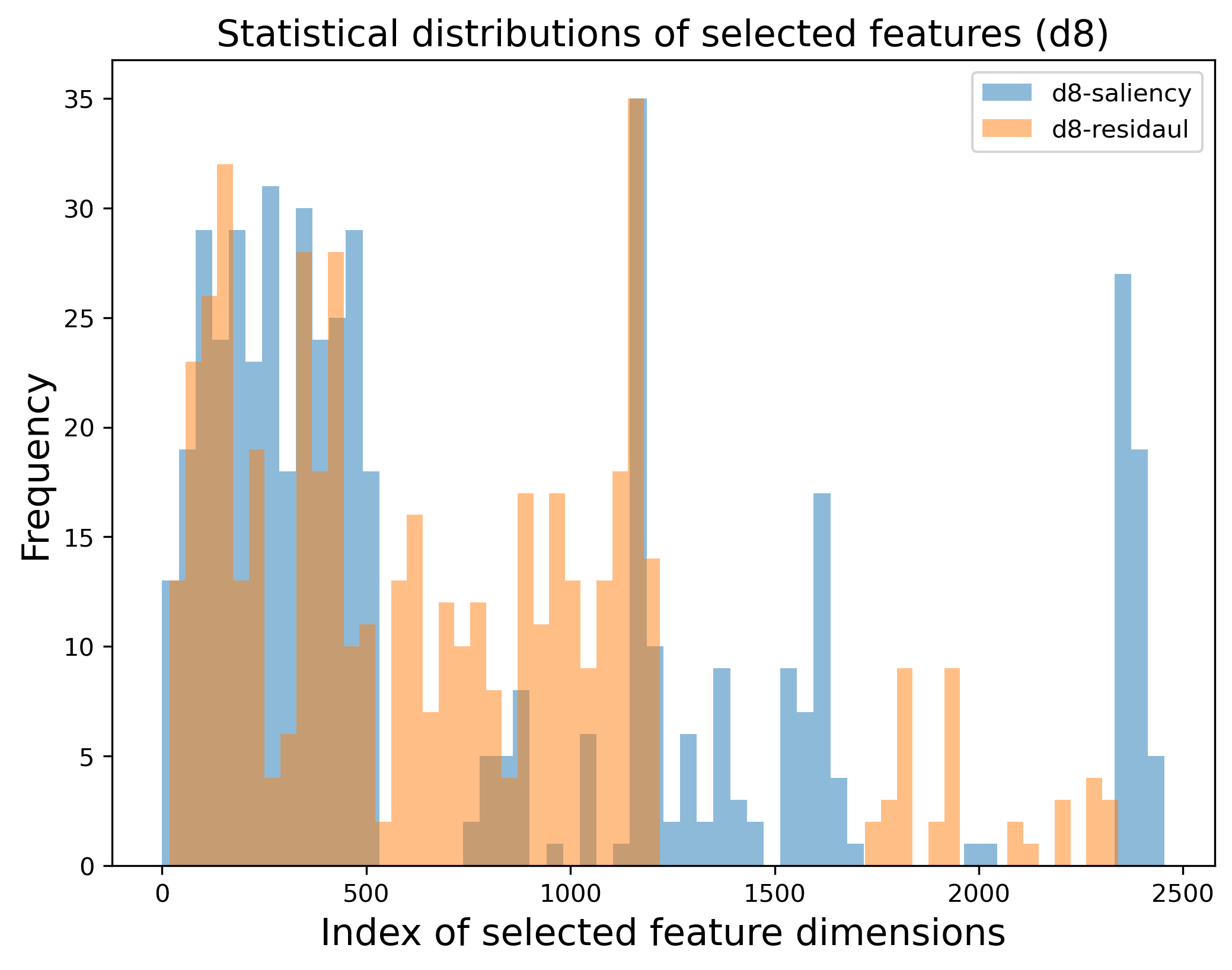}  \\
\scriptsize(a) \hspace{5cm} \scriptsize(b)
\caption{Distribution of RFT results in: (a) d16, and (b) d8 layers.}\label{fig:rft_distribution}
\end{figure*}

\subsubsection{Saliency Map Prediction and Saliency Residual Prediction}\label{subsubsec:residual}

In each path, leveraging the set of hybrid features extracted from the multi-layer hybrid feature extraction module, an initial step involves utilizing these features to predict their respective saliency maps.
For instance, in the path utilizing d64 features, the labels correspond to ground truth saliency maps down-sampled by a factor of 64. Subsequently, an XGBoost regressor is trained to perform pixel-wise prediction, mapping the feature domain to the saliency map domain.
Following the direct utilization of extracted hybrid features for saliency map prediction, the predicted saliency maps from d8 and d16 are transmitted to the ensemble module. In contrast, those from d32 and d64 undergo saliency residual prediction.
The saliency residual computation determines the disparity between the up-sampled predicted saliency map in the current layer and the ground truth within a shallower layer.
For instance, given the predicted saliency map in d64, the residual is calculated between the up-sampled d64 saliency map and the ground truth in d16.
Subsequently, d16 features train and predict this saliency residual using another regressor. The resultant predicted saliency residual in d16 is then added to the up-sampled predicted saliency map in d64 to form the final predicted saliency map originating from d64.

In summary, d64 and d32 features exclusively contribute to predicting their corresponding saliency maps, while d16 and d8 features predict both saliency maps and saliency residuals.
The rationale underlying the saliency residual prediction lies in the observation that d64 and d32 features primarily capture high-level features, resulting in the comparatively lower resolution of their predicted saliency maps.
Consequently, d8 and d32 features are leveraged to refine and calibrate the predicted saliency maps derived from d16 and d64, respectively. This strategy ensures enhanced precision and fidelity across multiple layers of predicted saliency maps.

Within the pipeline of multi-path saliency prediction, RFTs are employed to identify the most influential features for saliency map prediction and residual prediction.
It is noteworthy that, for a given set of features, distinct RFTs are utilized to select disparate subsets of features for saliency map prediction and saliency residual prediction.
This distinction arises due to the differing labels associated with these two predictions.
The label for the former prediction corresponds to the downsampled ground truth. In contrast, the label for the latter prediction represents the residual between the predicted saliency map and the ground truth.
The statistical distributions of the selected features employed for predicting saliency maps and saliency residuals on layers d16 and d8 are depicted in Figure \ref{fig:rft_distribution}.
Despite some overlapping observed between the two distributions, several distinct features are selected for the two types of predictions.
Given the label disparity between these two prediction tasks, selecting different features validates the necessity of employing separate RFTs for the same set of features.

\subsubsection{Ensembles}\label{subsubsec:ensemble}

An ensemble module is deployed to integrate the four predicted saliency maps spanning from d64 to d8. Within this module, the first step involves upsampling the four predicted saliency maps to match the resolution size of d4 and concatenating them. 
Subsequently, for each pixel location, the neighboring saliency values from the four saliency maps within a $5 \times 5$ block centered at this location are collated, yielding a 100-dimensional vector. 
Ultimately, an XGBoost regressor is trained using labels derived from downsampled ground truth by a factor of 4, facilitating the prediction of the saliency map in the d4 layer.

\subsubsection{Post-processing}\label{subsubsec:post}

Upon obtaining the fused predicted saliency maps in d4 and upsampling them to align with the resolution of the input images, several post-processing operations are implemented to furnish the final predicted saliency maps.
These operations entail three sequential steps. Initially, a portion of small saliency values is filtered out, as their presence after ensembling signifies a lack of confidence in prediction. Eliminating these small predicted saliency values aids in reducing false positive predictions.
Subsequently, a small Gaussian filter, with dimensions of $10 \times 10$, is applied to the entire image to enhance its smoothness. 
Finally, normalization is conducted on the entire image to align its distribution with the ground truth, thereby refining the accuracy and reliability of the predicted saliency maps.

\section{Experiments}\label{G_sec:experiments}

\subsection{Experimental Setup}\label{G_subsec:setup}

\subsubsection{Datasets}
Experiments were conducted utilizing the SALICON \cite{jiang2015salicon} and MIT300 \cite{judd2012benchmark} datasets, with benchmarks in the domain of image saliency detection.
The SALICON dataset, widely recognized within the research community, comprises a substantial collection of images, including 10,000 images in the training set, 5,000 in the validation set, and an additional 5,000 in the testing set.
On the other hand, the MIT300 dataset comprises 300 testing images, while the ground truth for this dataset is not publicly available.
To mitigate this limitation, the model was trained on the MIT1003 \cite{judd2009learning} dataset and subsequently evaluated on the MIT300 dataset, a practice commonly adopted by various benchmarks.
The MIT1003 dataset, consisting of 1,003 images, features ground truth annotations obtained through eye-tracking devices from 15 observers. 
Similarly, the MIT300 dataset was constructed following comparable procedures, drawing from the same image repositories and annotation methodologies.

\subsubsection{Evaluation Metrics}
To evaluate the performance of our saliency model comprehensively, we employ widely recognized metrics, including the linear correlation coefficient (CC), area under the receiver operating characteristic curve (AUC-J), shuffled AUC (s-AUC), normalized scanpath saliency (NSS), and similarity (SIM).
These metrics offer multifaceted insights into model efficacy in predicting eye fixation patterns. 
Notably, higher values of CC, AUC, sAUC, NSS, and SIM signify superior performance of the saliency model. 
For a more detailed explanation of these metrics and their relevance to saliency prediction, a comprehensive study is conducted in \cite{kummerer2018saliency}.

\subsubsection{Implementation Details}
The initial step in our experimental setup involves resizing the input images to dimensions of $(480 \times 640) \times 3$, utilizing the YUV color format.
Subsequently, we implement the multi-layer hybrid feature extraction process. Specifically, after computing the Saab coefficients from Saab $3 \times 3$ in layers d8, d16, and d32, the RFTs are employed to select 20, 50, and 100 coefficients, respectively, from each layer, which are then forwarded to the subsequent layer. 
A similar approach is applied to the Saab $5 \times 5$ coefficients.
In the multi-path saliency prediction phase, RFTs select various features from different layers: 500 from d8 features, 500 from d16 features, 1,000 from d32 features, and 1,000 from d64 features, respectively.
Regarding the training and testing partitioning, we adhere to the official splitting protocol for the SALICON dataset. For the MIT1003 dataset, we allocate 900 images for training and 103 for validation. Subsequently, testing is conducted on the 300 images comprising the MIT300 dataset.
All experiments are executed on a server equipped with an Intel(R) Xeon(R) E5-2620 CPU, ensuring consistency and reliability across computations.

\subsection{Experimental Results}
\subsubsection{Benchmarking Methods}\label{subsubsec:benchmarking}

We conducted a comprehensive performance evaluation of GreenSaliency compared to ten benchmarking methods, as summarized in Table \ref{table:saliency_MIT300} and Table \ref{table:saliency_SALICON}.
These benchmarking methods encompass conventional and DL-based image saliency detection methods, which we categorize into two groups for clarity.
\begin{itemize}
\item ITTI \cite{itti1998model} and GBVS \cite{harel2006graph}.
They are conventional image saliency detection methods that do not rely on neural networks.
\item Shallow Convnet \cite{pan2016shallow}, GazeGAN \cite{che2019gaze}, EML-NET \cite{jia2020eml}, Deep Convnet \cite{pan2016shallow}, SAM-ResNet \cite{cornia2018predicting}, SalFBNet \cite{ding2022salfbnet}, UNISAL \cite{droste2020unified}, and DeepGaze IIE \cite{linardos2021deepgaze}.
This category encompasses diverse DL-based image saliency detection methods, with transfer learning from outside datasets.
\end{itemize}

\begin{table}[t]
\centering
\caption{Performance comparison in five metrics between our GreenSaliency method and ten benchmarking methods on the MIT300 dataset.}
\begin{tabular}{ l c | c c c c c }
\hline
&Methods & AUC-J$\uparrow$ & s-AUC$\uparrow$  &CC$\uparrow$ & SIM$\uparrow$ & NSS$\uparrow$ \\
\hline
&ITTI \cite{itti1998model} &0.543 &0.535 &0.131 &0.338 &0.408 \\
&GBVS \cite{harel2006graph} &0.806 &0.630 &0.479 &0.484 &1.246\\
\hline
&Shallow Convnet \cite{pan2016shallow} &0.800 &0.640 &0.530 &0.460 &1.430 \\
&GazeGAN \cite{che2019gaze} &0.860 &0.731 &0.758 &0.649 &2.211 \\
&EML-NET \cite{jia2020eml} &0.876 &0.746 &0.789 &0.675 &2.487 \\
&Deep Convnet \cite{pan2016shallow} &0.830 &0.690 &0.580 &0.520 &1.510 \\
&SAM-ResNet \cite{cornia2018predicting} &0.852 &0.739 &0.689 &0.612 &2.062 \\
&SalFBNet \cite{ding2022salfbnet} &0.876 &0.785 &0.814 &0.693 &2.470 \\
&UNISAL \cite{droste2020unified} &0.877 &0.784 &0.785 &0.674 &2.369 \\
&DeepGaze IIE \cite{linardos2021deepgaze}  &\textbf{0.882} &\textbf{0.794} &\textbf{0.824} &\textbf{0.699} &\textbf{2.526} \\
\hline
&GreenSaliency &0.843 &0.700 &0.752 &0.647 &1.713 \\
\hline
\end{tabular}
\label{table:saliency_MIT300}
\end{table}

\begin{table}[t]
\centering
\caption{Performance comparison in five metrics between our GreenSaliency method and ten benchmarking methods on the SALICON dataset.}
\begin{tabular}{ l c | c c c c c }
\hline
&Methods & AUC-J$\uparrow$ & s-AUC$\uparrow$  &CC$\uparrow$ & SIM$\uparrow$ & NSS$\uparrow$ \\
\hline
&ITTI \cite{itti1998model} &0.667 &0.610 &0.205 &0.378 &- \\
&GBVS \cite{harel2006graph} &0.790 &0.630 &0.421 &0.446 &- \\
\hline
&Shallow Convnet \cite{pan2016shallow} &0.836 &0.670 &0.596 &0.520 &1.458 \\
&GazeGAN \cite{che2019gaze} &0.864 &0.736 &0.879 &0.773 &1.899 \\
&EML-NET \cite{jia2020eml} &0.866 &0.746 &0.868 &0.774 &\textbf{2.058} \\
&Deep Convnet \cite{pan2016shallow} &0.858 &0.724 &0.622 &0.609 &1.859 \\
&SAM-ResNet \cite{cornia2018predicting} &0.865 &0.741 &\textbf{0.899} &\textbf{0.793} &1.990 \\
&SalFBNet \cite{ding2022salfbnet} &0.868 &0.740 &0.892 &0.772 &1.952 \\
&UNISAL \cite{droste2020unified} &0.864 &0.739 &0.879 &0.775 &1.952 \\
&DeepGaze IIE \cite{linardos2021deepgaze}  &\textbf{0.869} &\textbf{0.767} &0.872 &0.733 &1.996 \\
\hline
&GreenSaliency &0.839 &0.679 &0.765 &0.683 &1.605 \\
\hline
\end{tabular}
\label{table:saliency_SALICON}
\end{table}

\subsubsection{Performance Evaluation}
We compare the performance of GreenSaliency with ten benchmarking methods in Table \ref{table:saliency_MIT300} and Table \ref{table:saliency_SALICON}.
GreenBaliency outperforms two conventional image saliency detection methods (i.e., ITTI and GBVS) and some earlier DL-based methods (i.e., Shallow Convnet and Deep Convnet) by a substantial margin in both two datasets.  
This shows the effectiveness of GreenSaliency in extracting multi-layer hybrid features to cover information from disparate receptive fields.  
GreenSaliency is also competitive with some DL-based methods (i.e., GazeGAN and SAM-ResNet). 
Compared with the state-of-the-art DL-based methods (i.e., EML-NET, SalFBNet, UNISAL, and DeepGAZE IIE), there is a gap to reach their performance. However, our model complexity is much lower than those, which is illustrated in Section \ref{subsec:complexity}.

\begin{figure}[!htbp]
\centering
\includegraphics[width=1.0\linewidth]{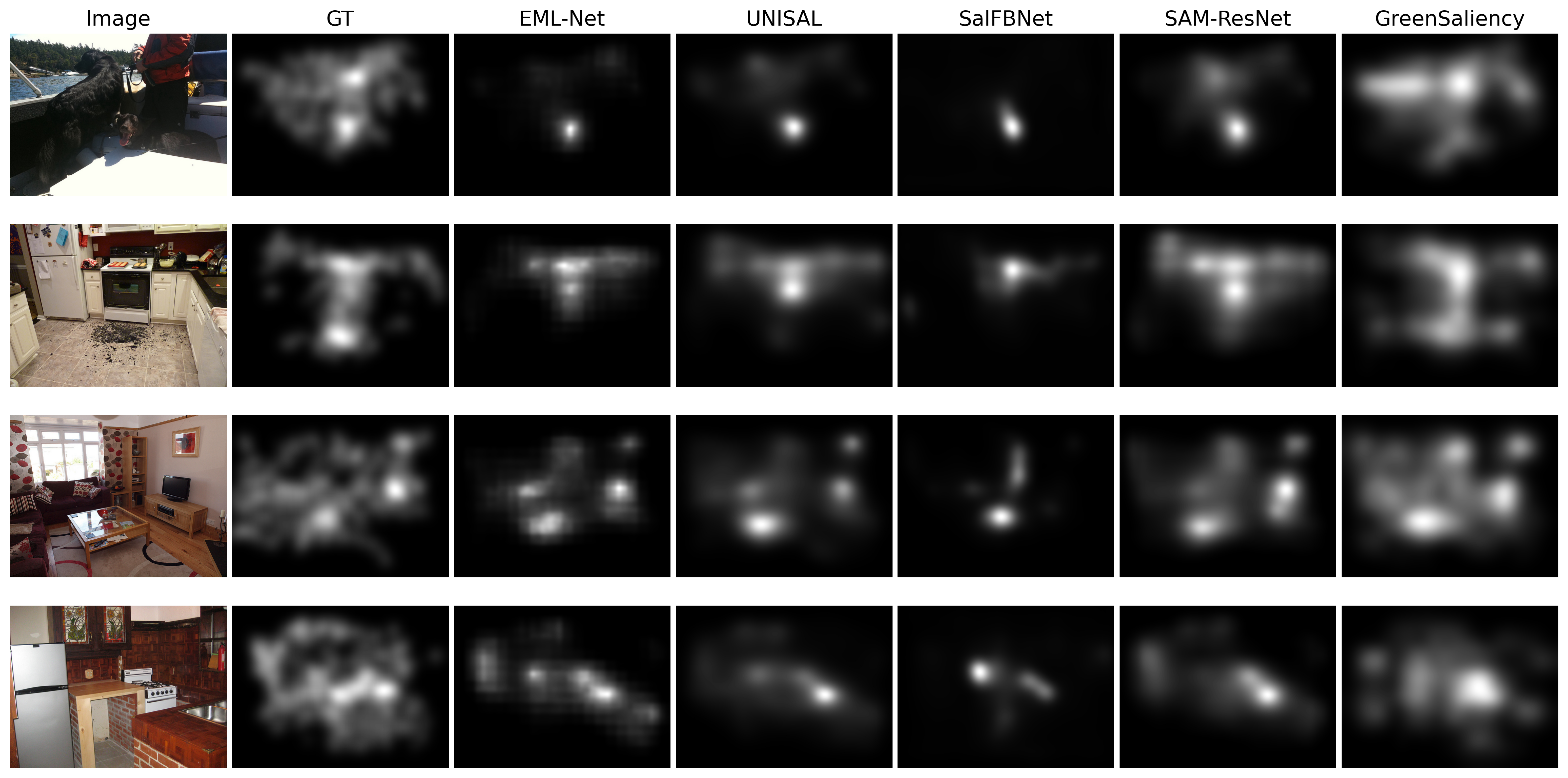}\\
\caption{Successful cases in GreenSaliency.}\label{fig:good_case}
\end{figure}

\begin{figure}[!htbp]
\centering
\includegraphics[width=1.0\linewidth]{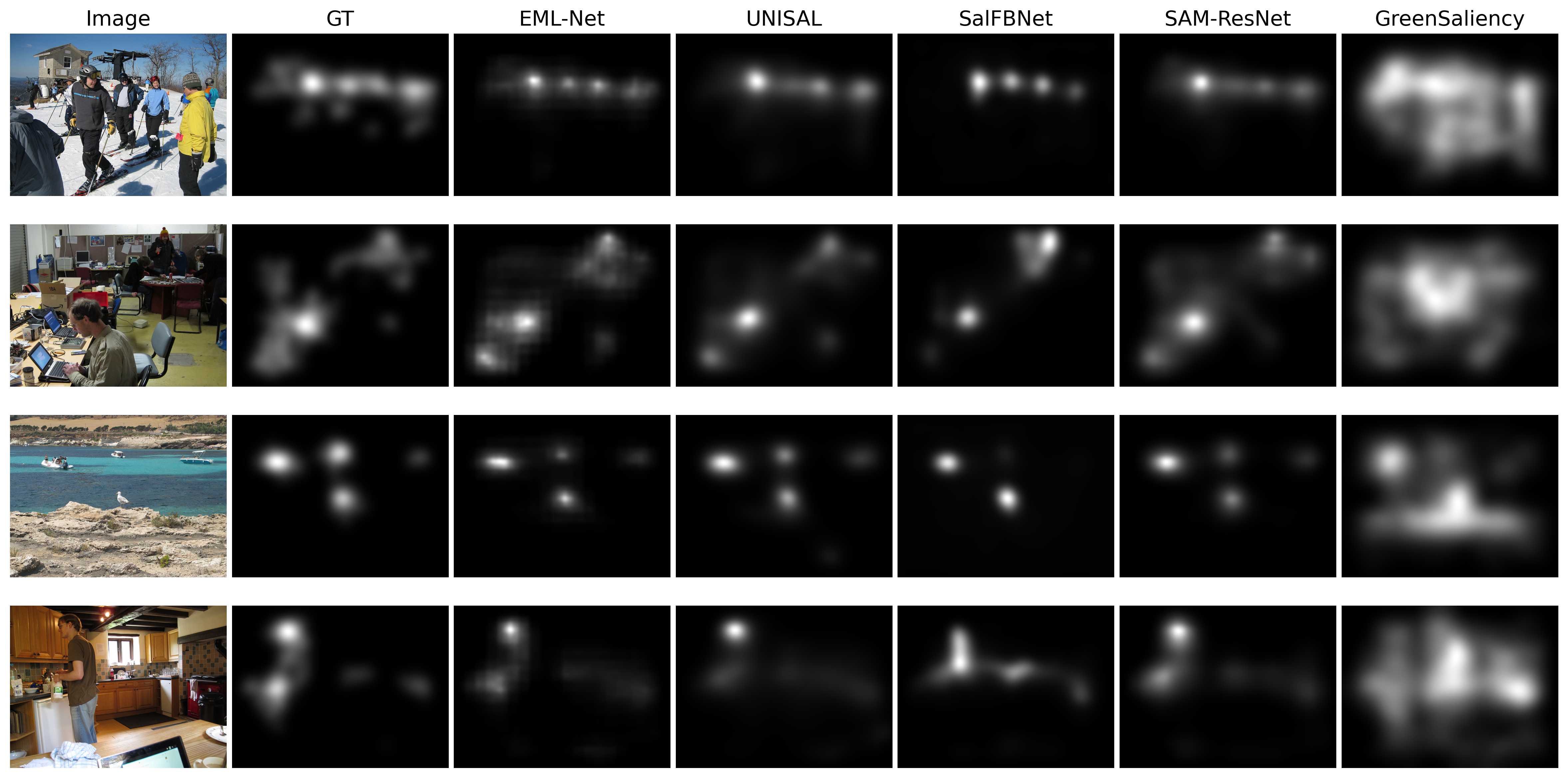}\\
\caption{Failed cases in GreenSaliency.}\label{fig:bad_case}
\end{figure}

Additionally, we conducted a qualitative analysis by comparing the predicted saliency maps generated by GreenSaliency with those produced by four benchmark methods.
Exemplary images showcase instances where GreenSaliency excels in Figure \ref{fig:good_case}.
Notably, the ground truth saliency map typically exhibits a dispersed and smoothed distribution in images featuring multiple objects without a dominant focal point.
Leveraging its approach, GreenSaliency demonstrates proficiency in attending to all objects within an image without relying on transfer learning from disparate datasets, such as ImageNet, for image classification.
Consequently, GreenSaliency performs relatively well on such images.
Conversely, in scenarios where images prominently feature highly captivating objects, such as humans and animals, as depicted in Figure \ref{fig:bad_case}, benchmark methods tend to prioritize these salient entities, particularly human faces.
Our proposed GreenSaliency, lacking in transfer learning tailored to specific object categories, may exhibit suboptimal performance in such contexts.

\begin{table*}[t]
\centering
\caption{Comparison of no. of model parameters,
model sizes (memory usage), no. of GigaFlops, and latency time of several saliency detection methods tested on the SALICON dataset, where ``X'' denotes the multiplier compared to our proposed method.}\label{table:complexity}
\resizebox{\linewidth}{!}{
\begin{tabular}{ l c | c c c c} \hline
&Model & \#Params (M)$\downarrow$ &Model Size (MB)$\downarrow$ &GFLOPs$\downarrow$ &Runtime (s)$\downarrow$\\ \hline
&Deep Convnet \cite{pan2016shallow} &25.5 (37.5X) &99 (34.7X) &3.2 (20.0X) &0.412 (10.8X) \\
&GazeGAN \cite{che2019gaze} &208.9 (307.2X) &879.2 (308.5X) &25 (156.2X) &2.540 (66.8X) \\
&EML-NET \cite{jia2020eml} &43 (63.2X) &180.2 (63.2X) &9.8 (61.3X) &0.365 (9.6X) \\
&Shallow Convnet \cite{pan2016shallow} &620 (911.7X) &2500 (877.2X) &3.9 (24.4X) &0.672 (17.7X) \\
&SAM-ResNet \cite{cornia2018predicting} &128.8 (189.4X) &535 (187.7X) &18.8 (117.5X) &0.858 (22.6X)\\
&SalFBNet \cite{ding2022salfbnet} &5.9 (8.7X) &23.4 (8.2X) &2.30 (14.4X) &0.180 (4.7X) \\
&UNISAL \cite{droste2020unified} &3.8 (5.6X) &14.7 (5.2X) &1.98 (12.4X) &0.083 (2.2X) \\
&DeepGaze IIE \cite{linardos2021deepgaze}  &98.2 (144.4X) &401 (140.7X) &21.2 (132.5X) &6.436 (169.4X) \\
\hline
&GreenSaliency &\textbf{0.68 (1X)} &\textbf{2.85 (1X)} &\textbf{0.16 (1X)} &\textbf{0.038 (1X)} \\
\hline
\end{tabular}}
\end{table*}

\subsection{Model Complexity}\label{subsec:complexity}
The significance of a lightweight model in saliency detection cannot be overstated, especially when it functions as a preliminary processing component in various computer vision applications. 
Moreover, the complexity of the model plays a pivotal role in determining its suitability for deployment on mobile and edge devices.
Our analysis assesses the model complexity of saliency detection methods across three key dimensions: model sizes, inference time, and computational complexity measured in terms of floating-point operations (FLOPs).
These metrics are presented comprehensively in Table \ref{table:complexity}, offering insights into different saliency detection approaches' efficiency and practical feasibility.

\subsubsection{Model Sizes}

The size of a learning model can be assessed through two primary metrics: 1) the total number of model parameters and 2) the actual memory usage. Model parameters can be represented in either floating-point or integer format, typically occupying 4 bytes and 2 bytes of memory, respectively.
Given that most model parameters are in floating point, the actual memory usage can be estimated as approximately four times the number of model parameters (as depicted in Table \ref{table:complexity}).
For clarity, we utilize the term ``model size'' to denote memory usage throughout the subsequent discussion. The model sizes of GreenSaliency and eight DL-based benchmark methods are detailed in the second column of Table \ref{table:complexity}.
Notably, GreenSaliency exhibits significantly lower model size than the two DL-based methods featuring lightweight models (i.e., SalFBNet and UNISAL).
In contrast to other DL-based methods characterized by notably large model sizes (often exceeding 100MB), GreenSaliency's model size ranges from 34 times to 877 times smaller than these counterparts.

\subsubsection{Inference Time}

An essential metric for assessing computational efficiency in image saliency detection is the inference time required to generate a saliency map.
Our comparative analysis evaluated the inference time of various DL-based methods on a server equipped with an Intel(R) Xeon(R) E5-2620 CPU. 
The inference time for predicting a single saliency map is documented in the fourth column of Table \ref{table:complexity}.
Notably, GreenSaliency demonstrates a significantly reduced inference time compared to other DL-based methods. Specifically, GreenSaliency achieves an inference time of 0.038 seconds per saliency map prediction, translating to an approximate processing speed of 26 frames per second, utilizing solely CPU resources.
It is imperative to acknowledge that as a non-DL-based method, GreenSaliency may not leverage computing acceleration resources as extensively as DL-based counterparts.
Nevertheless, with foreseeable advancements in third-party libraries and coding optimizations, GreenSaliency can realize even more excellent efficiency benefits in CPU or GPU-supported environments.

\subsubsection{Computational Complexity}

The assessment of computational complexity in saliency detection methods can be further elucidated by considering the number of floating-point operations (FLOPs) required.
To this end, we estimated the FLOPs for several DL-based methods essential for predicting a saliency map, compared with those of GreenSaliency.
The ``GFLOPs'' column in Table \ref{table:complexity} presents the number of GFLOPs necessary to execute a model once to generate a saliency map.
In line with our inference time analysis, GreenSaliency exhibits notably lower computational complexity than other DL-based methods.
Expressly, GreenSaliency necessitates 0.16 GFLOPs to predict a single image, whereas other DL-based methodologies require over 2 GFLOPs, resulting in a reduction ranging from 12 to 156 times lower in computational complexity.



\begin{table*}[!htbp]
\centering
\caption{Ablation study for GreenSaliency on SALICON dataset.}

\begin{tabular}{l l|c c c c c}
\hline
&Layer & AUC-J & s-AUC  &CC & SIM & NSS\\\hline
&d64 &0.822 &0.635 &0.710 &0.616 &1.428\\
&d64 + RP &0.827 &0.651 &0.734 &0.655 &1.505\\
\hline
&d32 &0.829 &0.639 &0.721 &0.611 &1.425\\ 
&d32 + RP &0.834 &0.658 &0.748 &0.661 &1.516\\ 
\hline
&d16 &0.831 &0.660 &0.726 &0.628 &1.425\\
\hline
&d8 &0.830 &0.662 &0.725 &0.635 &1.428\\
\hline
&ensmeble &0.839 &0.679 &0.765 &0.683 &1.605\\
\hline
\end{tabular}
\label{table:ablation}
\end{table*}

\subsection{Ablation Study}\label{subsec:ablation}

To evaluate the individual contributions of various components to the overall performance of GreenSaliency, we conducted an ablation study as outlined in Table \ref{table:ablation}.
This study assessed the impact of different feature sets, namely d64, d32, d16, and d8, along with saliency residual prediction (RP) as depicted in Figure \ref{fig:multi-path}.
Specifically, we independently investigated the efficacy of each set of hybrid features for layers d64, d32, d16, and d8. Our findings revealed that employing a single set of features resulted in the highest performance for the d16 layer, whereas the d64 layer exhibited the lowest performance.
Upon incorporating residual prediction, notable enhancements in performance were observed for the d64 and d32 layers, underscoring the importance of residual prediction in improving overall performance.
Moreover, through ensembling the performance metrics across all four layers, we observed further improvements across all five evaluation metrics, culminating in achieving the highest performance values.

\section{Conclusion and Future Work}\label{G_sec:conclusion}
This paper introduces a novel lightweight image saliency detection approach named GreenSaliency, which operates without using DNNs or pre-training on external datasets. 
GreenSaliency surpasses all conventional (non-DL-based) image saliency detection methods and achieves comparable performance with some early-stage DL-based methods. 
Compared to state-of-the-art DL-based methods, GreenSaliency exhibits lower prediction accuracy but offers advantages in smaller model sizes, shorter inference time, and reduced computational complexity. 
The minimal model complexity of GreenSaliency suggests a more efficient energy usage, making it suitable for integration into extensive image processing systems.

Our proposed GreenSaliency method identifies several limitations, which could be addressed in future iterations. Notably, GreenSaliency struggles to accurately predict particular popular objects of interest, such as human faces and animals, while tending to emphasize all objects within an image, as illustrated in Figure \ref{fig:bad_case}. 
Consequently, implementing a mechanism to rank the saliency levels of different objects could enhance prediction accuracy by focusing on the most appealing objects. 
Additionally, given the significance of saliency detection in various computer vision tasks, such as blind image quality assessment (BIQA), there exists potential and interest in developing a saliency-guided approach for addressing BIQA challenges.

\bibliographystyle{unsrt}  
\bibliography{references}

\end{document}